\documentclass[pdftex,twocolumn,epjc3]{svjour3}          

\RequirePackage[T1]{fontenc}

\smartqed  
\RequirePackage{graphicx}
\RequirePackage{mathptmx}      
\RequirePackage{flushend}
\RequirePackage[numbers,sort&compress]{natbib}
\RequirePackage[colorlinks,citecolor=blue,urlcolor=blue,linkcolor=blue]{hyperref}
\RequirePackage{epstopdf}
\journalname{Eur. Phys. J. C}
\RequirePackage{amsmath,amssymb}
\begin{document}

\title{Dark photon dark matter and fast radio bursts}


\author{Ricardo G. Landim\thanksref{e1,addr1}
}

\thankstext{e1}{ricardo.landim@tum.de}

\institute{Physik Department T70, James-Franck-Stra$\beta$e 1,
Technische Universit\"at M\"unchen, 85748 Garching, Germany\label{addr1}
}

\date{Received: date / Accepted: date}

\maketitle

\begin{abstract}
The nature of dark matter (DM) is still a mystery that may indicate the necessity for extensions of the Standard Model (SM). Light dark photons (DP) may comprise partially or entirely the observed DM density and existing limits on the DP DM parameter space arise from several cosmological and astrophysical sources. In the present work we investigate DP DM using cosmic transients, specifically fast radio bursts (FRBs). The observed time delay of radio photons with different energies have been used to constrain the photon mass or the Weak Equivalence Principle, for example. Due to the mixing between the visible and the DP, the time delay of photons from these cosmic transients, caused by free electrons in the intergalactic medium, can change and impact those constraints from FRBs. We use five detected FRBs and two associations of FRBs with gamma-ray bursts to investigate the correspondent variation on the time delay caused by the presence of DP DM. The result is virtually independent of the FRB used and this variation is very small, considering the still allowed  DP DM parameter space, not jeopardizing current bounds on other contributions of the observed time delay.
  \end{abstract}

\section{Introduction}
Dark matter (DM) is one the biggest puzzles in cosmology, and lately in particles physics, whose existence is a hint for physics beyond the standard model (SM). Weakly Interacting Massive Particles (WIMPs) have been the most well-known  DM
candidates (see Ref.\cite{Arcadi:2017kky} for a review), but the lack of positive signatures opens new avenues of exploration. Among several extensions of the SM, a new $U(1) $ gauge field was proposed as mediator of the interaction between DM and SM particles \cite{Feldman:2006wd,Feldman:2007wj,Pospelov:2007mp,Pospelov:2008zw,Davoudiasl:2012qa,Davoudiasl:2012ag,Essig:2013lka,Izaguirre:2015yja, Belotsky:2004ga,Fargion:2005ep,Khlopov:2006uv,Belotsky:2006fd,Belotsky:2006pp,Khlopov:2006dk}. The so-called `dark photon' (DP) interacts with the visible sector through the kinetic-mixing with the hypercharge field\footnote{We use the same notation of Ref. \cite{Dubovsky:2015cca}.} $\frac{\varepsilon}{2\sqrt{1+\varepsilon^2}} F_{\mu\nu}F'^{\mu\nu}$ \cite{Holdom:1985ag,Holdom:1986eq,Dienes:1996zr,DelAguila:1993px,Babu:1996vt,Rizzo:1998ut,Khlopov:2002gg,Caputo:2020bdy,Caputo:2020rnx}. Its  parameter space has been constrained through plenty of observations and experiments  \cite{Aartsen:2012kia,Aartsen:2016exj,Battaglieri:2017aum,Riordan:1987aw,Bjorken:1988as,Bross:1989mp,Bjorken:2009mm, Pospelov:2008zw,Davoudiasl:2012ig,Endo:2012hp,Babusci:2012cr,Archilli:2011zc,Adlarson:2013eza,Abrahamyan:2011gv,Merkel:2011ze,Reece:2009un,Kazanas:2014mca,Chang:2016ntp,Cardoso:2017kgn,Brito:2015oca,Baryakhtar:2017ngi,Bauer:2018onh,Cardoso:2018tly,Caputo:2019tms}, and has led to theoretical explanations for the (expected) smallness of the kinetic-mixing parameter \cite{Dienes:1996zr,Abel:2003ue,Abel:2006qt,Landim:2019epv, Landim:2019gdj,Gherghetta:2019coi,Koren:2019iuv}.  

One of the DM candidates is a very light DP, whose viable production mechanisms have been investigated over the years \cite{Karciauskas:2010as,Himmetoglu:2009qi,Himmetoglu:2008zp,Nelson:2011sf,Arias:2012az,Graham:2015rva,Agrawal:2018vin,Agrawal:2018vin,Dror:2018pdh,Co:2018lka,Bastero-Gil:2018uel,Nakayama:2019rhg,AlonsoAlvarez:2019cgw,Long:2019lwl,Nakai:2020cfw}. Several cosmological and astrophysical constraints are applied to DP DM, arising from the on-shell and off-shell (resonant and non-resonant) transition between DP and
SM photons \cite{Arias:2012az,McDermott:2019lch, Witte:2020rvb}, the non-resonant absorption of DP and subsequent heating of  Milky Way’s
interstellar medium \cite{Dubovsky:2015cca}, the heating/cooling of the Leo T dwarf galaxy \cite{Wadekar:2019xnf}, the heating of cold gas clouds \cite{Bhoonah:2018gjb}, the Ly-$\alpha$ forest \cite{Irsic:2017yje}, CMB spectral distortions \cite{McDermott:2019lch,Garcia:2020qrp}, the heating of the intergalactic medium (IGM) at the epoch of helium reionization, and the depletion of DM and energy deposition during dark ages \cite{McDermott:2019lch}.

A very interesting astrophysical phenomenon still to be understood, but that has various application in cosmology and astrophysics is   fast radio burst (FRB). FRBs are very bright and brief ($\sim$ ms) cosmic transients of unknown origin, discovered in 2007 \cite{Lorimer:2007qn} (see \cite{Pen:2018ilo,Platts:2018hiy, Popov:2018hkz, Petroff:2019tty} for  recent reviews). Among the  explanations for its origin are  mergers and interactions between compact objects, supernovae remnants and active galactic nuclei \cite{Platts:2018hiy}.  Regardless of its possible source, FRBs have been used, for instance, to constrain  cosmological parameters \cite{Zhou:2014yta,Yang:2016zbm,Walters:2017afr,Li:2017mek,Yu:2017beg,Wei:2018cgd,Zitrin:2018let},  to study the properties of the  IGM \cite{Deng:2013aga,Zhang:2013lta,Akahori:2016ami,Fujita:2016yve,Shull:2017eow,Ravi:2018ose}, compact DM \cite{Munoz:2016tmg,Wang:2018ydd},  to set upper limits on the photon mass \cite{Xing:2019geq,Wu2016,Wei:2016jgc,Bonetti:2016cpo,Bonetti:2017pym, Wei:2018pyh,Shao:2017tuu}, and to constrain  the Weak Equivalence Principle (WEP) \cite{Wei:2015hwd,Tingay:2015wdv,Shao2017,Yu:2017xbb,Bertolami:2017opd,Yu:2018slt,Xing:2019geq,Wang:2019tjh,LeviSaid:2020mbb}.

Since the visible photon mixes with DP, the dispersion velocity of the former changes when compared with the absence of the mixing. This fact is reflected in the photon frequency, when it travels through the IGM, and as a result the time delay caused by the dispersion in the IGM $\Delta t_{IGM}$ may change. In turn, this change could influence current bounds on the photon mass or WEP, for instance.

In this paper we apply five detected FRBs (FRB  110220, FRB 121102, FRB 150418, FRB 180924 and FRB  190523) and two association of FRBs with gamma-ray bursts (GRB) (FRB/GRB 101011A and FRB/GRB 100704A) to investigate the contribution of DP DM to the time delay caused by IGM effects. The seven FRBs have their source measured or inferred in a redshift range of $0.15<z<1$, whose time delays of photons with different frequencies (mostly between 1.2 GHz and 1.5 GHz) are in the range $0.15$ s $\lesssim \Delta t_{obs}\lesssim 1$ s. The variation of the time delay caused by IGM effects is very small for the still allowed DP DM parameter space, not jeopardizing current bounds on other contributions of the observed time delay.



We organize the paper in the following manner. Sect. \ref{sec:DP} reviews the necessary expressions for DP DM in a charged plasma. In Sect. \ref{sec:FRB} we present the detected FRBs and the resulting variation of the IGM time delay. Sect. \ref{sec:conclu} is reserved for conclusions.

\section{Dark photon in a plasma}\label{sec:DP}

After diagonalizing the  photon and DP kinetic terms, the DP Lagrangian takes the form
\begin{equation}
    \mathcal{L}\supset -\frac{1}{4}F_{\mu\nu}'F'^{\mu\nu}+\frac{m_{A'}^2}{2}A'^{\mu}A'_\mu-\frac{e}{(1+\varepsilon^2)}J_\mu(A^\mu+\varepsilon A'^\mu)\,,
\end{equation}
where $J_\mu$ is the SM electric current.

The IGM has a plasma  frequency given by\footnote{This expressions does not take into account the influence of inhomogeneities on the electron number density. It was explored in \cite{Witte:2020rvb}, but does not influence our results.}
\begin{equation}
    \omega_p(z)=\sqrt{\frac{4 \pi \alpha n_e(z)}{m_e}}\,,
\end{equation}
where $\alpha$ is the fine-structure constant, $m_e$ is the electron mass and $n_e(z)=n_{e,0}(1+z)^3$ is the free electron number density, where the electron number density today is $n_{e,0}\sim 10^{-7}$ cm$^{-3}$ \cite{Garcia:2020qrp}. 

The mixing between  visible and  hidden photon  changes the photon  dispersion relation. In order to reach the appropriate expression, the Proca and Maxwell equations are solved along with the  equations for a non-relativistic plasma \cite{Dubovsky:2015cca}. In the case of a non-relativistic DP, which is assumed as a DM candidate  ($k\ll \omega$), the longitudinal and transverse components of the gauge fields obey the same (mixed) equation. The diagonalization of this equation gives \cite{Dubovsky:2015cca}
\begin{align}\label{eq:photonomega}
  \omega^2_\pm&= \frac{1}{2}\Bigg\{m_{A'}^2+\frac{\omega_p^2}{1+i \frac{\nu}{\omega}}\pm \bigg[\bigg(m_{A'}^2+\frac{\omega_p^2}{1+i \frac{\nu}{\omega}} \bigg)^2\nonumber\\& -\frac{4\omega_p^2 m_{A'}^2}{(1+\varepsilon^2)(1+i \frac{\nu}{\omega})}\bigg]^{1/2}\Bigg\}\,,
\end{align}
where $\nu$ is the frequency of electron-ion collisions
\begin{equation}
    \nu=\frac{4\sqrt{2\pi}\alpha^2n_e}{3m_e^{1/2}T_e^{3/2}}\log\left(\frac{4\pi T_e^3}{\alpha^3 n_e}\right)^{1/2}.
\end{equation}
The electron temperature is about $T_e\sim 10^4-10^7$ K in the IGM \cite{Meiksin:2007rz}, and as we shall see $\omega\sim \omega _p$, thus $\nu\ll \omega$ for the parameters in our range of interest. Therefore, for the purpose of this work,  we may neglect the imaginary part of Eq. (\ref{eq:photonomega}). 

From Eq. (\ref {eq:photonomega}) we have  $\omega_+\geq \omega_p$, while $\omega_-\leq\omega_p$. When $m_{A'}^2\ll \omega_p^2$ Eq. (\ref {eq:photonomega}) becomes
\begin{align}\label{eq:omegagammaless}
    \omega_\gamma^2=\omega_+^2&= \omega_p^2+\frac{\varepsilon^2 m_{A'}^2}{1+\varepsilon^2} +\mathcal{O}\left(m_{A'}^3\right)\,,\\
  \omega_{A'}^2&  =\omega_-^2=\frac{ m_{A'}^2}{1+\varepsilon^2} +\mathcal{O}\left(m_{A'}^3\right)\,,
\end{align}
where $\omega_\gamma$ is the photon frequency, while $\omega_{A'}$ is the DP frequency. On the other hand, when $m^2_{A'} \gg \omega_p^2$ the frequencies are
\begin{align}\label{eq:omegagammagreater}
    \omega_\gamma^2=\omega_-^2&= \frac{ \omega_p^2}{1+\varepsilon^2} +\mathcal{O}\left(\omega_p^4\right)\,,\\
      \omega_{A'}^2=\omega_+^2&= m_{A'}^2+\frac{\varepsilon^2 \omega_p^2}{1+\varepsilon^2} +\mathcal{O}\left(\omega_p^4\right)\,.
\end{align}
The positive-sign solution in Eq. (\ref{eq:photonomega}) behaves as the photon for $m_{A'}^2< \omega_p^2$, while for $m_{A'}^2> \omega_p^2$ is the negative-sign solution that represents the photon frequency.  

We show in Fig. \ref{figexcludedregions} the existing limits on the DP DM parameter space. 

\section{Fast radio bursts and dark photon dark matter}\label{sec:FRB}

Due to the interaction between photon and DP, the frequency of the former when it travels through the IGM is no longer $\omega_p$, but given by Eq. (\ref{eq:photonomega}). 

The observed time delay  $\Delta t_{obs}$ for FRB photons with different energies may have the following contributions \cite{Gao:2015lca,Wei:2015hwd}
\begin{equation}\label{tobs}
\Delta t_{obs}=\Delta t_{int}+\Delta t_{LIV}+\Delta t_{spe}+\Delta t_{DM}+\Delta t_{grav}\, ,
\end{equation}
where $\Delta t_{int}$ is the intrinsic astrophysical time delay, $\Delta t_{LIV}$ represents the time delay due to (possible) Lorentz invariance violation, $\Delta t_{spe}$ is  the time delay caused by  photons with a non-zero rest mass,   $\Delta t_{grav}$ is the  Shapiro time delay and $\Delta t_{DM}$ is the time delay from the dispersion by the line-of-sight free electron content. 

We are interested only in the time delay due to dispersion by free electrons, thus we can ignore all other sources of delay. This time delay $\Delta t_{DM}$, in turn, has contributions due to the host galaxy, the IGM and the Milky Way. However, the host galaxy is usually unknown and the contribution from the Milky Way is much smaller than the one from the IGM \cite{Cordes:2002wz,prochaska2019probing}, so that we can consider the the limit where the dispersion measure time delay is solely due to the IGM $\Delta t_{DM}\approx \Delta t_{IGM}$. This limit is translated to conservative bounds on the other contributions in Eq. (\ref{tobs}), as constraints on the photon mass \cite{Wu:2016brq} or the WEP \cite{Wei:2015hwd}, which can be even more constrained if other contributions are  taken into account. 

The IGM magnetic effect on the dispersion velocity of  photons can be ignored because the Larmor frequency is much smaller than the plasma frequency. The time delay due to the IGM plasma on two  photons with frequencies $\nu_l$ and $\nu_h$ is \cite{Wu:2016brq}
\begin{equation}\label{eq:tigm}
   \Delta t_{\omega_\gamma}=\frac{\nu_{\gamma,0}^2}{2 H_0}(\nu^{-2}_l-\nu^{-2}_h)H_2(z)\,,
\end{equation}
where $\nu_{\gamma,0}=\omega_{\gamma,0}/(2\pi )$,
\begin{equation}
    H_2(z)=\int_0^z\frac{(1+z')dz'}{\sqrt{\Omega_m(1+z')^3+\Omega_\Lambda}}\,,
\end{equation}
and we will adopt the cosmological parameters from the \textit{Planck} satellite, $\Omega_m=0.315$, $\Omega_\Lambda=0.685$, $H_0= 100h$ km s$^{-1}$ Mpc$^{-1}$, and $h=0.674$ \cite{Aghanim:2018eyx}.


The correspondent variation in the time delay caused by IGM when DP DM is present is
\begin{equation}\label{eq:DP}
\Delta t_{DP}=\Delta t_{\omega_\gamma}-\Delta t_{IGM}\,,
\end{equation}
where $\Delta t_{IGM}$ is the time delay when $\omega_\gamma=\omega_p$. 

In order obtain $\Delta t_{DP}$, we use five detected FRBs and two combinations of FRBs and gamma-ray bursts (GRB):\footnote{Further information and a list of all detected FRBs can be found at http://www.frbcat.org/.}
\begin{itemize}
    \item FRB 110220 was discovered by  the 64-m Parkes telescope \cite{Thornton:2013iua}, localized to
coordinates (J2000) RA = 22h 34m, Dec $= -12^{\circ}24'$ for photons ranging between 1.2 GHz and 1.5 GHz, and whose inferred redshift of 0.81 was estimated through its dispersion measure value. 
    \item GRB 101011A was detected by \textit{Swift}/BAT  with coordinates (J2000) RA = 03h13m12s, Dec $= -65^{\circ}$ $59'
08''$ \cite{Cannizzo2010}, while   GRB 100704A was detected by  the \textit{Fermi} Gamma-Ray Burst Monitor \cite{McBreen2010} and \textit{Swift}/BAT \cite{Grupe2010} with coordinates (J2000)
RA = 08h54m33s
, Dec = $-24^{\circ}$ $12'
55''$. The association systems FRB/GRB  were observed between the frequencies 1.23 GHz and 1.45 GHz \cite{bannister2012limits}, and had the redshift estimated using the Amati relation \cite{Deng:2013aga}.

    \item FRB 121102 was the first repeating FRB observed, in the Arecibo PALFA pulsar survey  with coordinates (J2000)  RA = 05h31m58s, Dec $=+33^{\circ}08'04''$ \cite{Spitler:2014fla}, in the frequency range 1.23 -- 1.53 GHz, and is one the few FRBs that has its redshift precisely determined at $z=0.193$ \cite{Chatterjee:2017dqg,Tendulkar:2017vuq,Bassa:2017tke,Marcote:2017wan}. 
    
    \item FRB 150418 was also detected by the Parkes telescope \cite{Keane:2016yyk} in the frequency range 1.2 -- 1.5 GHz. Although its redshift was claimed to be measured \cite{Keane:2016yyk}, its localization was contested \cite{Williams:2016zys}. More recently its redshift was  constrained \cite{Walker:2018qmw} and the result is similar to the original claim.
    \item FRB 180924 was detected by the ASKAP telescope   at J2000 coordinates RA = 21h44m25.255s
and Dec =  $-40^{\circ}$ $ 54'00''
$, between frequencies 1.2 GHz and 1.5 GHz \cite{Bannister:2019iju}. The redshift of the host galaxy was determined  as $z=0.32$.
    \item  FRB 190523 was detected by the Deep
Synoptic Array ten-antenna prototype (between frequencies 1.35 GHz and 1.50 GHz) \cite{Ravi:2019alc}  and was localized to
J2000 coordinates RA = 13h48m15.6s and Dec $= +72^{\circ}28'11''$. Its redshift of 0.66 is also one of the few that were determined.
\end{itemize}    

The correspondent observed time delay, the time delay caused by IGM effect for the frequency $\omega_\gamma=\omega_p$, and the  (inferred or measured) redshift are shown in Table \ref{table1}. 

\begin{figure*}
    \centering
    \includegraphics[scale=0.65]{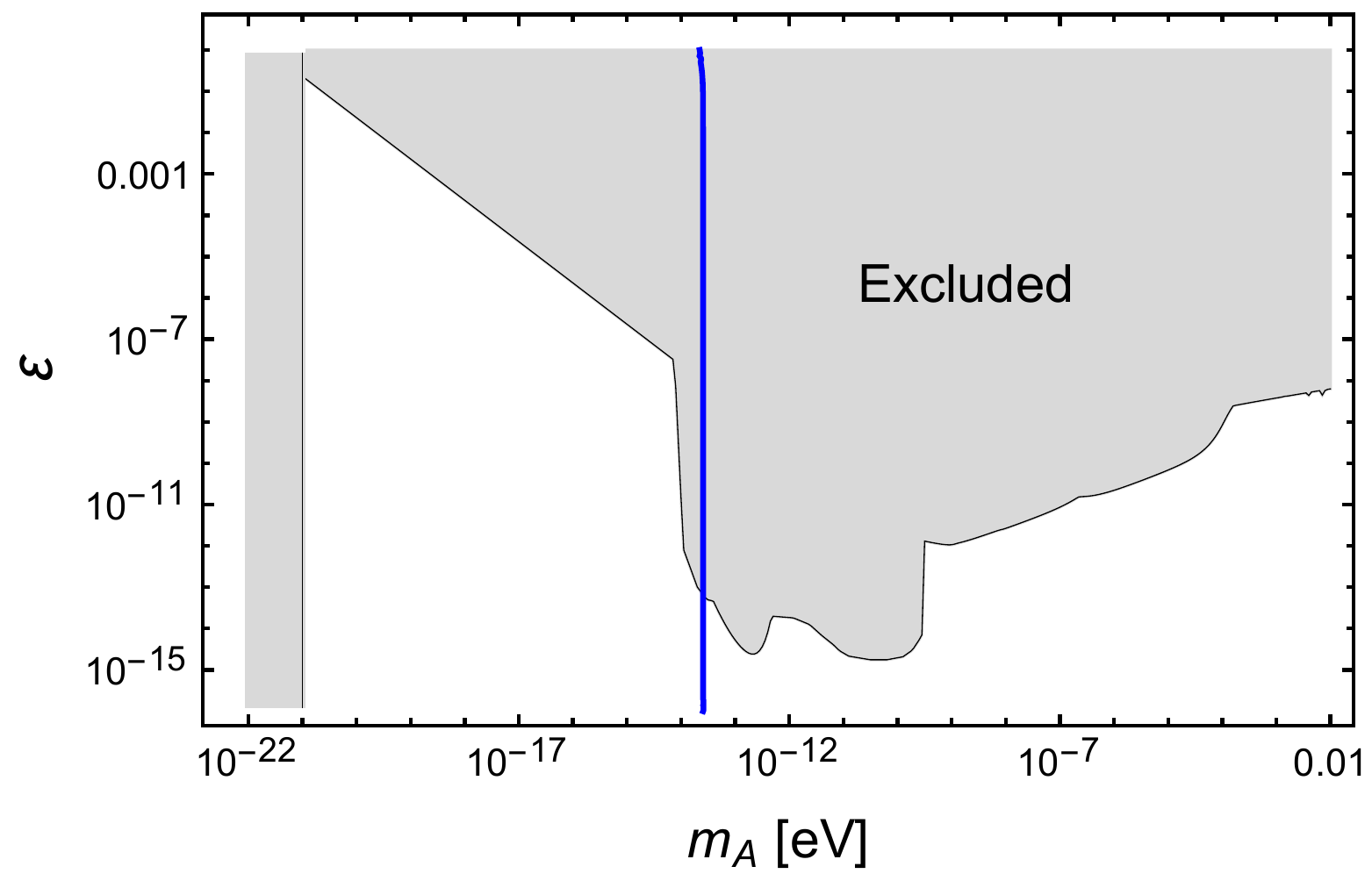}
    \caption{Existing limits on DP DM presented in \cite{McDermott:2019lch,Witte:2020rvb}, which include several cosmological and astrophysical constraints \cite{McDermott:2019lch,Witte:2020rvb,Arias:2012az,Dubovsky:2015cca,An:2013yfc,Redondo:2013lna,Vinyoles_2015,Irsic:2017yje,Bhoonah:2018gjb,Wadekar:2019xnf}, where the vertical blue line is the plasma frequency today. }
    \label{figexcludedregions}
\end{figure*}

\begin{figure*}
    \centering
    \includegraphics[scale=0.5]{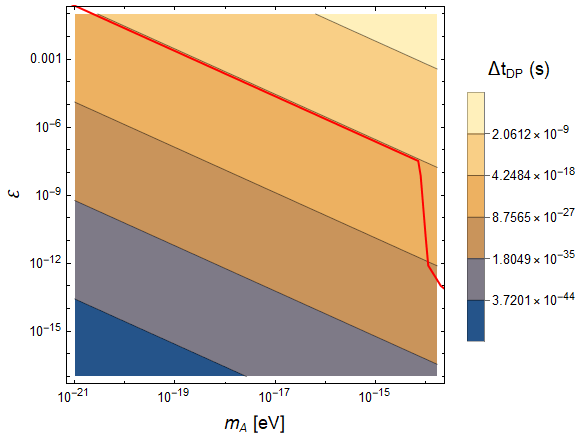}
    \caption{Extra time delay $\Delta t_{DP}$ (\ref{eq:DP})  due to DP DM for the region $m_{A'}\leq\omega_p$. We used the FRB 121102 ($\nu_l=1.23$ GHz, $\nu_h=1.53$ GHz and $z=0.193$), however all other FRBs present virtually the same behavior.  It is shown in red the existing limits, as in Fig. \ref{figexcludedregions}.}
    \label{figure2}
\end{figure*}

\begin{figure*}
    \centering
    \includegraphics[scale=0.5]{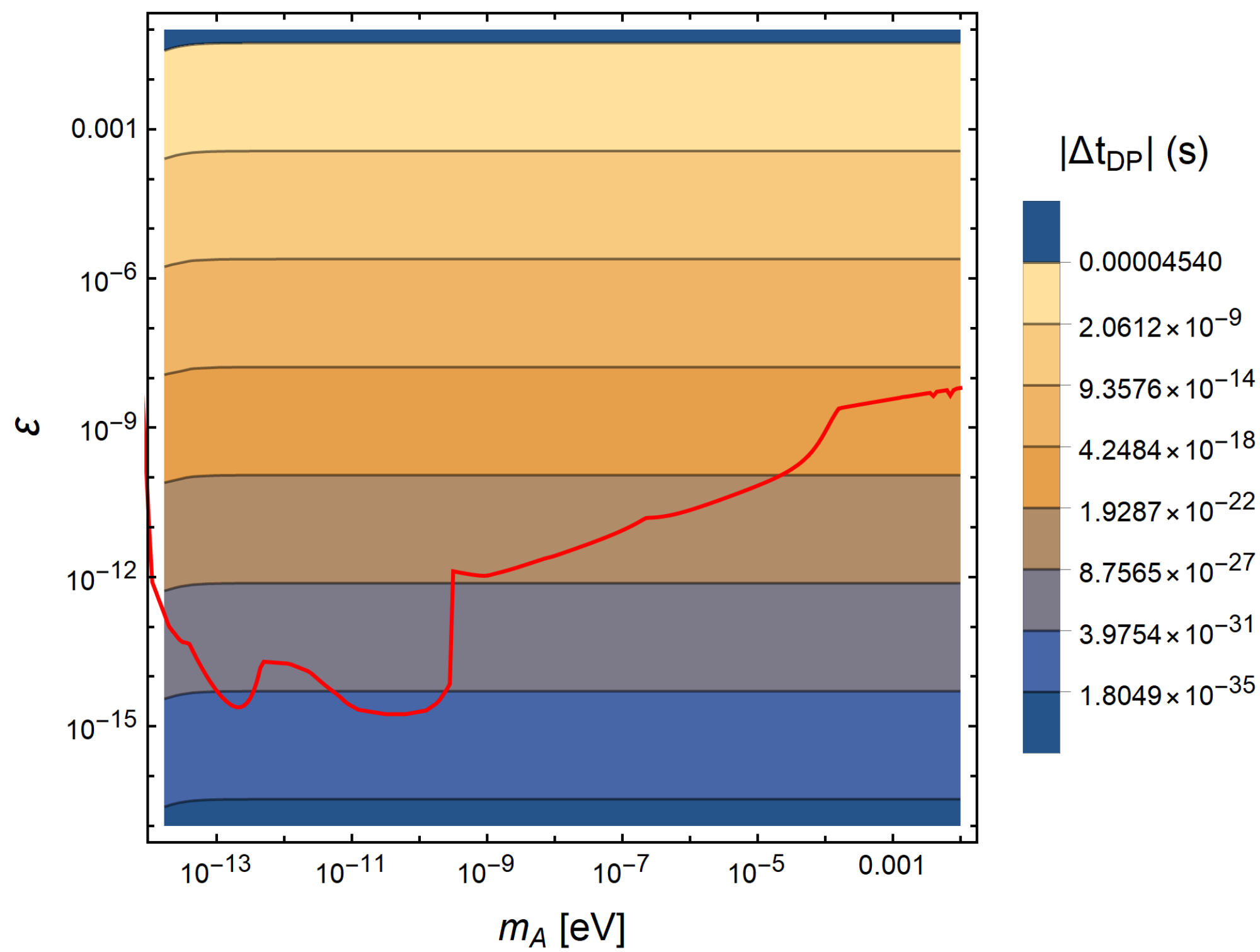}
    \caption{Extra time delay $\Delta t_{DP}$ (\ref{eq:DP})  due to DP DM for the region $m_{A'}\geq\omega_p$.  We used the FRB 121102 ($\nu_l=1.23$ GHz, $\nu_h=1.53$ GHz and $z=0.193$), however all other FRBs present virtually the same behavior. It is shown in red the existing limits, as in Fig. \ref{figexcludedregions}.}
    \label{figure3}
\end{figure*}

    Using the seven FRBs, we apply Eqs. (\ref{eq:photonomega}) and (\ref{eq:DP}) to obtain the contribution $\Delta t_{DP}$, whose results are presented in Figs. \ref{figure2} and \ref{figure3}. The observed time delay may follow this hierarchical relation  $\Delta t_{obs}\geq \Delta t_{\omega_\gamma}\geq \Delta t_{IGM}$, where only the positive branch $\omega_+$ can reach  a value that would give an increase in $\Delta t_{IGM}$, possible only for $m_{A'}<\omega_p$. On the other hand, for $m_{A'}>\omega_p$ the extra time delay is $\Delta t_{DP}\approx -\varepsilon^2 \Delta t_{IGM}$, which in turn  is much smaller than the one for the ultra-light DP region, considering the still allowed parameter space in Fig. \ref{figexcludedregions}. The results are very similar for all FRBs used and we plot only one of them (FRB 121102). The extra time delay $\Delta t_{DP}$ is very small to have a considerable influence on Eq. (\ref{tobs}), therefore it does not impact the existing bounds from the other terms in  $\Delta t_{obs}$.

    

   

\begin{table}[t]
\begin{tabular}{ c c c c }
\hline
	 & $z$ & $\Delta t_{obs}$ (s) & $\Delta t_{IGM}$ (s)  \\
\hline
FRB 110220  &  0.81  	  &  1    &  0.90      \\
FRB 121102  &  0.193  	  &  0.55     &  0.19      \\
FRB 150418 &  0.492  	  &  0.815     &   0.54     \\
FRB 180924  &  0.32  	  &  0.40     &  0.36     \\
FRB 190523  &  0.66  	  &  0.33     & 0.29       \\
 FRB/GRB 101011A &  0.246 &  0.438  & 0.19     \\
 FRB/GRB 100704A  &    0.166    &     0.149     & 0.13     \\
    \hline
\end{tabular}
\caption{Redshift, observed time delay and time delay $\Delta t_{IGM}$,  corresponding to Eq. (\ref{eq:tigm}) for $\omega_{\gamma}=\omega_{p}$, for seven detected FRBs.  }\label{table1}
\end{table}

\section{Conclusions}\label{sec:conclu}

In this paper we have investigated light DP DM using seven detected FRBs. These observations lie in the redshift range $0.1<z<0.8$ and have observed time delays between $0.1 \text{ s}<\Delta t_{obs}\leq 1$ s. Due to the mixing between the photon and DP, the photon dispersion relation is no longer equal to the plasma frequency of the IGM, but depends on the DP mass and the kinetic-mixing parameter.   Therefore, taking the conservative scenario where the time delay between  radio photons of different frequencies  caused by their dispersion through the electron plasma is solely due to the IGM, we obtained the possible contribution to this time delay from DP DM.  The results are practically insensitive to the FRBs, and the corresponding extra time delays $\Delta t_{DP}$ are very small to have an impact in current observations. 

The extra time delay $\Delta t_{obs}-\Delta t_{IGM}$ of FRB photons has been used to constrain other contributions in Eq. (\ref{tobs}). Therefore, DP DM does not give a considerable variation in $\Delta t_{IGM}$, which otherwise could change existing  limits on photon mass or the WEP, for instance.

On the other hand, decreasing the error associated with the determination  of the dispersion measure (as it can be seen in some FRB data from www.frbcat.org) and improving our knowledge about FRBs, such as their origin,  will in turn be translated to an improvement on the determination of the dispersion measure from IGM. Therefore, it is expected that, in the future, better constraints on the dispersion measure will provide a reliable form to obtain better limits on the DP DM parameter space.

\begin{acknowledgements}
 We thank Samuel Witte and Gonzalo  Alonso-Álvarez for comments. This work was supported by CAPES (process 88881.162206/2017-01) and Alexander von Humboldt Foundation.\end{acknowledgements}

\bibliographystyle{unsrt}
\bibliography{references}\end{document}